# Ultralow Voltage Operation of p- and n-FETs Enabled by Self-Formed Gate Dielectric and Metal Contacts on 2D Tellurium

Chang Niu[1,2,†], Linjia Long[1,2,†], Yizhi Zhang[3], Zehao Lin[1,2], Pukun Tan[1,2], Jian-Yu Lin[1,2], Wenzhuo Wu[4], Haiyan Wang[3] and Peide D. Ye[1,2,*]

[1]*Elmore Family School of Electrical and Computer Engineering, Purdue University, West Lafayette, IN 47907, United States.*

[2]*Birck Nanotechnology Center, Purdue University, West Lafayette, IN 47907, United States.*

[3]*School of Materials Science and Engineering, Purdue University, West Lafayette, Indiana 47907, United States.*

[4]*School of Industrial Engineering, Purdue University, West Lafayette, IN 47907, United States.*

†These authors contributed equally to this work: Chang Niu, Linjia Long

*Correspondence and requests for materials should be addressed to P. D. Y. (yep@purdue.edu)

The ongoing demand for more energy-efficient, high-performance electronics is driving the exploration of innovative materials and device architectures, where interfaces play a crucial role due to the continuous downscaling of device dimensions. Tellurium (Te), in its two-dimensional (2D) form, offers significant potential due to its high carrier mobility and ambipolar characteristics, with the carrier type easily tunable via surface modulation. In this study, we leverage atomically controlled material transformations in 2D Te to create intimate junctions, enabling near-ideal field-effect transistors (FETs) for both n-type and p-type operation. A $NiTe_x$-Te contact provides highly transparent interfaces, resulting in low contact resistance, while the $TiO_x$-Te gate dielectric forms an ultraclean interface with a capacitance equivalent to 0.88 nm equivalent oxide thickness (EOT), where the quantum capacitance of Te is observed. Subthreshold slopes (SS) approach the Boltzmann limit, with a record-low SS of 3.5 mV/dec achieved at 10 K. Furthermore, we demonstrate 2D Te-based complementary metal-oxide-semiconductor (CMOS) inverters operating at an ultralow voltage of 0.08 V with a voltage gain of 7.1 V/V. This work presents a promising approach to forming intimate dielectric/semiconductor and metal/semiconductor junctions for next-generation low-power electronic devices.

Modern computational technology is experiencing a significant surge in power consumption driven by the increasing density and speed of electronic devices. As integrated circuits continue to evolve, the demand for higher performance pushes the limits of device scaling, leading to greater power dissipation and thermal challenges.[1] Among various approaches to enhance energy efficiency, scaling down the supply voltage $V_{DD}$ has emerged as the most effective strategy, given that dynamic power consumption is proportional to the square of the supply voltage. This means even modest reductions in $V_{DD}$ can lead to substantial power savings, making voltage scaling a key focus in the design of low-power electronics. However, reducing $V_{DD}$ presents significant challenges, especially as complementary metal-oxide-semiconductor (CMOS) field-effect transistors (FETs) continue to downscale to atomic level. At these scales, the quality of interfaces between metal/semiconductor[2–11] and dielectric/semiconductor,[12–17] becomes critical in determining device performance.

In this paper, we present a simple and effective approach to achieving atomically sharp and ultraclean interfaces in 2D Te by precisely controlling chemical reactions and material transformations at the atomic scale. We demonstrate near-ideal n-type and p-type MOSFETs using 2D Te, featuring ultralow operation voltage CMOS inverters down to 0.08 V, negligible hysteresis, a gate capacitance equivalent to 0.88 nm equivalent oxide thickness (EOT), and a subthreshold slope approaching the Boltzmann limit in a wide temperature range. This study highlights the potential of 2D Te for next-generation low-power and cryogenic electronic applications, offering a pathway for reducing power consumption and enhancing device performance through advanced interface engineering.

**2D Te FETs with self-formed metal contacts and gate dielectrics**

A 40 nm layer of Ti and Ni was directly deposited on top of 2D Te surface using electron-beam evaporation to form the top-gate structure, as illustrated in Figure 1a. A 90 nm $SiO_2$/Si substrate was used as the back gate. 2D Te FETs with various gate lengths were fabricated, as shown in the scanning electron microscopy (SEM) image in Figure 2b. To characterize the material composition and device structure, high-angle annular dark-field scanning transmission electron microscopy (HAADF-STEM) was performed. Figure 1c shows the cross-sectional HAADF-STEM image of a 2D Te FET device with Ni and Ti contacts, where a distinct contrast is observed beneath the Ni electrode, suggesting material chemical reaction and transformation occurred in 2D Te. To further investigate the difference between the Ni-Te and Ti-Te interfaces, energy dispersive x-ray spectroscopy (EDS) elemental mappings were conducted, as shown in Figure 1d and 1e. The line profiles across the interfaces reveal elemental distributions at the nanometer scale, clearly identifying the interfaces. The Ni-Te interface shows a significant Ni signal within the Te region, indicating the formation of a Ni-Te compound. In contrast, the Ti signal decreases rapidly to zero within the Te region, suggesting minimal diffusion of Ti atoms. An oxygen peak is observed at the Ti interface, where the Te signal is absent, which is ascribed to the formation of a $TiO_x$ layer. The thickness of this $TiO_x$ layer is estimated to be approximately 6 nm, based on the half-width of the peak. The formation of $TiO_x$ can be attributed to the reduction of a thin layer of native $TeO_x$ during the direct deposition of Ti. According to the Ellingham diagram,[18,19] which describes the Gibbs free energy ($\Delta G$) of metal oxide formation, Ti is more reactive than Te at room temperature and has a stronger tendency to form oxides. This thermodynamic preference results in the formation of an insulating $TiO_x$ layer that serves as a dielectric when Ti is deposited onto Te. Meanwhile, the process eliminates the native $TeO_x$ which usually provides the significant interface traps.

Figure 1f and 1g show HAADF-STEM images of 2D Te under Ti gate and Ni contact regions, respectively. Te possesses a unique chiral crystal structure, where three covalently bonded Te atoms in each unit cell form atomic chains along the c-direction. These atomic chains are arranged into a hexagonal lattice through van der Waals interactions. The single-crystalline Te structure with lattice constants of a = 4.5 Å and c = 6 Å, and a single atomic

Te chain is highlighted in Figure 1f. The crystal orientation corresponds to the (1000) facet (y-z plane), as illustrated in Figure 1h. In contrast, 2D Te under the Ni contact undergoes a significant material transformation, displaying a completely different crystal structure from Te. A layered $NiTe_x$ structure with an interlayer spacing of 7 Å is observed, with the lattice constant of $NiTe_x$ varying depending on the material composition. The formation of $NiTe_x$ can be controlled by adjusting the annealing temperature and time after Ni deposition. Figure 1i shows the electrical properties of different metal electrodes on 2D Te. Similar to Ti, the Te-Al interface is also insulating due to the formation of $Al_2O_3$, making Ti and Al suitable for gate materials. The metallic nature of $NiTe_x$ makes it an excellent intimate contact[20] for semiconducting 2D Te, similar to contact doping in 2D materials[21,22] and silicides in Si-based technologies.[23] Figure S1 presents the HAADF-STEM image at the channel-to-contact region, revealing a sub-1 nm atomically sharp $NiTe_x$-Te interface. The contact length ($L_c$) corresponds to the thickness of the 2D Te layer. The specific contact resistivity ($\rho_c$) of p-type 2D Te FET is extracted to be $8.3 \times 10^{-8}$ $\Omega$ $cm^2$, which is very low among p-type FETs,[24,25] indicating that a clean and transparent semiconductor-to-metal interface is formed.

**Device performance of 2D Te FETs and effects of quantum capacitance**

Equivalent-oxide-thickness (EOT) is one of the important parameters of the dielectric layer in MOSFETs. Scaling EOT down to sub-one-nanometer region[13,26–29] with a low leakage current is critical to have a good electrostatic control of the semiconducting channel and a lower operating voltage. Here using the self-formed $TiO_x$ as dielectric layer, we achieved an ultralow EOT of 0.88 nm and observed the quantum capacitance effect in 2D Te.

Figures 2a and 2b show the color maps of drain current ($I_d$) as a function of top-gate (self-formed $TiO_x$) and back-gate (90 nm $SiO_2$) voltages for a 2D Te FET in n-type and p-type regions, respectively, at a temperature of 10 K. The 2D Te thickness is around 6 nm.

Due to its narrow bandgap of 0.35 eV,[30] the carrier type in 2D Te can be effectively tuned via electrical gating. The transfer curves are presented in Figure S2a. Comparing the drain current responses, the top-gate is approximately 100 times more efficient than the back-gate in transconductance; specifically, a gate voltage change of 0.1 V at the top-gate yields the same current level as 10 V change at the back-gate. Notably, there is a difference in efficiency between the p-type and n-type regions. It is important to note that the top-gate and back-gate are not identical in terms of interface conditions. The back-gate interacts through van der Waals forces, while the top-gate involves a deposition process. Additionally, the contact region is influenced by the back gate, with a connecting region between the contact and the top-gate. Another representative device exhibiting similar behavior and top-gate tunability is shown in Figure S2.

Using the same device structure, the capacitance of top gate was measured as a function of top-gate and back-gate voltages, with the Ni electrode grounded, for both n-type and p-type 2D Te FET, as shown in Figure 2c and 2d, respectively. The capacitance maps exhibit similar behavior to the drain current maps, displaying distinct depletion and accumulation regions. The capacitance-equivalent-thickness (CET) reflects the tunability of the dielectrics integrated with the semiconductor. Typically, CET is higher than EOT due to the quantum capacitance ($C_q$) effect.[31] In conventional Si-based MOSFETs, the quantum capacitance is relatively large, making the total gate capacitance close to the oxide capacitance ($C_{ox}$). However, quantum capacitance effects become significant when materials exhibit unique DOS versus energy dispersion, as seen in graphene and carbon nanotubes,[32,33] and when the oxide capacitance is sufficiently large. In 2D Te, the narrow bandgap allows access to the transport properties of both the conduction and valence bands within the same device, making quantum capacitance effects observable and distinguishable, since 2D Te has high mobility or low effective mass of carriers in both sides. Figure 2e illustrates the band structure of 2D Te near the Fermi level. The conduction band minimum and valence band maximum are located around the H (H') point with two-fold valley degeneracy. Due to the three-fold screw symmetry and the strong spin-orbit coupling, the spin degeneracy of the conduction band splits and crosses at the H (H') point, forming a Weyl node.[34] The valence band also exhibits spin non-degeneracy with a camel-back structure. The quantum capacitance is directly proportional to the DOS and, under a parabolic approximation, can be expressed as:

$$C_q = \rho e^2 = \frac{g_s g_v m^* e^2}{2\pi \hbar^2}$$

where $\rho$ is the density of states, $g_s$ and $g_v$ is the spin and valley degeneracies, respectively, and $m^*$ is the effective mass. For valence band, $g_{sp} = 1$, $g_{vp} = 2$, and $m_p^* = 0.26 m_0$,[35] resulting in $C_{qp} = 17.52\ \mu F/cm^2$. For conduction band, $g_{sn} = 2$, $g_{vn} = 2$, and $m_n^* = 0.10 m_0$,[34,36] resulting in $C_{qn} = 13.48\ \mu F/cm^2$. It is noteworthy that at lower carrier densities, due to the spin-orbit coupling, the band deviates from a parabolic shape near the band edge,[37] making the parabolic approximation valid only at high carrier densities. The CET for p-type and n-type regions at high densities is extracted to be 1.06 nm and 1.14 nm, respectively, as shown in Figure 2f. Considering the quantum capacitance effect, the EOT of the same gate is calculated to be 0.88 nm for both p-type and n-type regions, providing strong evidence for the presence of quantum capacitance effects. The observed increase in capacitance at higher carrier densities further supports the spin-orbit interaction-induced quantum capacitance effect.

The self-formed Ti gate can also function as a back-gate where the TiO$_x$ is formed during the sample fabrication, mitigating limitations associated with the resistance of the contact and the contact-to-gate connecting region. The threshold voltage (V$_t$) is controlled by growing a thin layer of Al$_2$O$_3$ (1 to 3 nm) on the Te surface,[38] effectively removing the intrinsic doping effects from the native tellurium oxide layer. Utilizing highly transparent metal contacts and efficient gate dielectrics, the electrical performance of the self-formed 2D Te n-FET and p-FET devices, each with a channel length of 1 µm at 33 K, is presented in Figure 3a,3b and Figure 3c,3d, respectively. Both n- and p-type devices exhibit an on/off current ratio of $10^6$ at $V_{ds}$ = 0.05 V, demonstrating excellent switching behavior. An on-current of 11 µA/µm is achieved with well-defined saturation behavior at $V_{ds}$ = 0.5 V and $V_{gs}$ = 0.5 V, demonstrating the potential for low-voltage $V_{DD}$ of 0.5 V operation. Moreover, the devices exhibit ultralow hysteresis of 4 mV and 2 mV at the $10^{-5}$ µA/µm current level in n-FET and p-FET configurations, respectively. This low hysteresis reflects the presence of a clean dielectric-to-semiconductor interface, resulting a low density of interfacial traps and enhanced device stability.

**Ultralow voltage operation of 2D Te CMOS inverters enabled by the ultraclean interfaces**

The ultraclean dielectric-to-semiconductor interface enables improved control of switching in the subthreshold region. Figure 4a presents the transfer curves of a 1 µm channel length device at $V_{ds}$ = 0.01 V and 0.05 V at 10 K. The transistors exhibit a steeper slope in the subthreshold region, with the device undergoing a 6-order-of-magnitude change in resistance within a 0.2 V gate voltage window. The subthreshold slope (SS) is calculated using the relation $SS = \partial log(I_d)/\partial V_{gs}$ at different current levels, as shown in Figure 4b. An ultralow SS of 3.5 mV/dec is achieved. Figure 4c and 4d display the temperature dependence of the transfer characteristics for the 2D Te n-FET and p-FET, respectively, from 10 K to 295 K. The extracted SS values are summarized in Figure 4e. The subthreshold slope can be expressed as: $SS = \ln(10)\frac{nkT}{q}, n = 1 + \frac{C_d+C_{it}}{C_{ox}}$, where $k$ is the Boltzmann constant, $q$ is the elementary charge, $T$ is temperature, $C_d$ is depletion layer capacitance, $C_{it}$ is interfacial trap capacitance, $C_{ox}$ is oxide capacitance.[14] In 2D Te, the SS is close to the Boltzmann limit (n = 1) and scales linearly with temperature in both n-type and p-type transistors under 200 K, indicating low interfacial trap states and high gate oxide capacitance. At higher temperatures (above 200 K), the SS deviates from the Boltzmann limit due to the thermal activation of electrons and holes in p- and n-FETs, owning to the narrow bandgap of Te. Notably, the SS in 2D Te does not saturate as the temperature decreases, unlike Si-based FETs, which show saturation at cryogenic temperatures.[39] This makes 2D Te MOSFETs a promising candidate for cryogenic-temperature computing.

2D Te FETs can be modeled as two back-to-back Schottky diodes in the thermionic emission regime. The barrier height $\Phi_B$ is extracted from temperature-dependent measurements using: $I_d \approx A^* T^{1.5} \exp(-\frac{q\Phi_B}{kT})$, with $q\Phi_B = q\Phi_{B0} - \frac{V_{gs}-V_{FB}}{n}$ for $V_{gs} < V_{FB}$, where $A^*$ is the Richardson constant, $V_{FB}$ is flat band voltage.[7] The gate-dependent barrier height, shown in Figure 4f, is calculated from the slope of the Arrhenius plots (Figure S3). The n factor is extracted to be 1.05 for n-FET and 0.93 p-FET based on barrier heights at $V_{gs} < V_{FB}$, providing further evidence of the ultraclean semiconductor-to-dielectric interface. The Schottky barrier height is extracted to be 38.7 mV and 21.7 mV for n-type and p-type transistors, respectively, at the flat band condition. In contrast to most of the 2D materials[40] or Ge-based devices,[41] where contacts suffer from large Schottky barriers due to the Fermi level pinning effect at the semiconductor-to-metal interface, the NiTe$_x$-Te interface offers near transparent contacts for both electrons and holes.

The near-ideal n-FETs and p-FETs based on 2D Te pave an alternative path for scaling down V$_{DD}$ of CMOS technology, thereby reducing power consumption. Figure 5a shows the voltage transfer characteristics of a 2D Te CMOS inverter at 10 K for different V$_{DD}$ values. A common input and output electrode structure is employed with the threshold voltage controlled by ALD capping of Al$_2$O$_3$. The device demonstrates ultralow operating voltage with V$_{DD}$ = 0.08 V, exhibiting full-output-swing behavior. The inverter gain ($\partial V_{out}/\partial V_{in}$) is presented in Figure 5b, where a high gain of 7.1 V/V is achieved at V$_{DD}$ = 0.08 V. Butterfly curves (Figure S4) with two storage states represent the performance of two cross-coupled inverters, which serve as the fundamental building blocks for static-random-access memory (SRAM). The noise margin dependence on $V_{DD}$ is extracted using the largest possible square method and is plotted in Figure 5b. A significant noise margin of 75% is achieved, highlighting the potential for stable and reliable low-voltage operation in 2D Te-based CMOS technology.

**Conclusion**

The formation of intimate interfaces through self-formed materials via direct metal deposition on 2D Te provides a novel approach to achieving near-ideal n-type and p-type MOSFETs. This strategy enables ultralow operating voltages of 0.08 V and a voltage gain of 7.1 V/V in 2D Te CMOS inverters, facilitated by the transparent metal/semiconductor $NiTe_x$-Te contacts and ultraclean dielectric/semiconductor $TiO_x$-Te interfaces. The $TiO_x$-Te interfaces exhibit low interfacial trap states and an ultrahigh gate capacitance with an equivalent oxide thickness (EOT) of 0.88 nm. Our findings present a scalable pathway for significantly reducing power consumption in future CMOS technologies by precisely engineering metal-semiconductor and dielectric-semiconductor interfaces.

**References**

1. IEEE International Roadmap for Devices and Systems. https://irds.ieee.org/editions (2023).

2. Du, Y., Liu, H., Deng, Y. & Ye, P. D. Device perspective for black phosphorus field-effect transistors : *ACS Nano* **8**, 10035–10042 (2014).

3. Wu, H. *et al.* Germanium nMOSFETs with recessed channel and S/D: contact, scalability, interface, and drain current exceeding 1 A/mm. *IEEE Trans. Electron Devices* **62**, 1419–1426 (2015).

4. Gül, Ö. *et al.* Hard superconducting gap in InSb nanowires. *Nano Lett.* **17**, 2690–2696 (2017).

5. Ridderbos, J. *et al.* Hard superconducting gap and diffusion-induced superconductors in Ge-Si nanowires. *Nano Lett.* **20**, 122–130 (2020).

6. Sumita, K., Takeyasu, J., Toprasertpong, K., Takenaka, M. & Takagi, S. Low specific contact resistance between InAs/Ni-InAs evaluated by multi-sidewall TLM. *AIP Adv.* **13**, 055310 (2023).

7. Shen, P. C. *et al.* Ultralow contact resistance between semimetal and monolayer semiconductors. *Nature* **593**, 211–217 (2021).

8. Li, W. *et al.* Approaching the quantum limit in two-dimensional semiconductor contacts. *Nature* **613**, 274–279 (2023).

9. Niu, C. *et al.* Record-low metal to semiconductor contact resistance in atomic-layer-deposited $In_2O_3$ TFTs reaching the quantum limit. *IEEE Int. Electron Devices Meet. IEDM* 5–8 (2023).

10. Niu, C. *et al.* Surface accumulation induced negative Schottky barrier and ultralow contact resistance in atomic-layer-deposited $In_2O_3$ thin-film transistors. *IEEE Trans. Electron Devices* **71**, 3403–3410 (2024).

11. Allain, A., Kang, J., Banerjee, K. & Kis, A. Electrical contacts to two-dimensional semiconductors. *Nat. Mater.* **14**, 1195–1205 (2015).

12. Xuan, Y., Shen, T., Xu, M., Wu, Y. Q. & Ye, P. D. High-performance surface channel In-rich $In_{0.75}Ga_{0.25}As$ MOSFETs with ALD high-k as gate dielectric. *IEEE Int. Electron Devices Meet. IEDM* 371–374 (2008).

13. Zeng, D. *et al.* Single-crystalline metal-oxide dielectrics for top-gate 2D transistors. *Nature* **632**, 788-794 (2024).

14. Lyu, J.-S. A new method for extracting interface trap density in short-channel MOSFETs from substrate-bias-dependent Subthreshold Slopes. *ETRI J.* **15**, 10–25 (1993).

15. Ashman, C. R., Schwarz, K., Fo, C. J. & Blo, P. E. The interface between silicon and a high-k oxide ¨. *Nature* **427**, 53–56 (2004).

16. Delabie, A. *et al.* Effective electrical passivation of Ge(100) for high-k gate dielectric layers using germanium oxide. *Appl. Phys. Lett.* **91**, 082904 (2007).

17. Mootheri, V. *et al.* Interface admittance measurement and simulation of dual gated CVD $WS_2$ MOSCAPs: mapping the $D_{it}(E)$ profile. *Solid. State. Electron.* **183**, 108035 (2021).

18. Hasegawa, Masakatsu. "Ellingham diagram." In Treatise on Process Metallurgy, *Elsevier*, 507-516 (2014).

19. Kleykamp, H. The chemical state of the fission products in oxide fuels. *J. Nucl. Mater.* **131**, 221–246 (1985).

20. Jena, D., Banerjee, K. & Xing, G. H. Intimate contacts. *Nat. Mater.* **13**, 1076–1078 (2014).

21. Jiang, J., Xu, L., Qiu, C. & Peng, L. M. Ballistic two-dimensional InSe transistors. *Nature* **616**, 470–475 (2023).

22. Jiang, J. *et al.* Yttrium-doping-induced metallization of molybdenum disulfide for ohmic contacts in two-dimensional transistors. *Nat. Electron.* **7**, 545-556 (2024).

23. Gambino, J. P. & Colgan, E. G. Silicides and ohmic contacts. *Mater. Chem. Phys.* **52**, 99–146 (1998).

24. Jung, Y. *et al.* Transferred via contacts as a platform for ideal two-dimensional

transistors. *Nat. Electron.* **2**, 187-194 (2019).

25. Xie, J. *et al.* Low resistance contact to p-type Monolayer WSe$_2$. *Nano Lett.* **24**, 5937–5943 (2024).

26. Li, W. *et al.* Uniform and ultrathin high-κ gate dielectrics for two-dimensional electronic devices. *Nat. Electron.* **2**, 563–571 (2019).

27. Zhang, L. *et al.* Vertically grown metal nanosheets integrated with atomic-layer-deposited dielectrics for transistors with subnanometre capacitance-equivalent thicknesses. *Nat. Electron.* **7**, 662–670 (2024).

28. Shen, Y. *et al.* Two-dimensional-materials-based transistors using hexagonal boron nitride dielectrics and metal gate electrodes with high cohesive energy. *Nat. Electron.* (2024) doi:10.1038/s41928-024-01233-w.

29. Zhu, C. *et al.* Magnesium niobate as a high-κ gate dielectric for two-dimensional electronics. *Nat. Electron.* (2024) doi:10.1038/s41928-024-01245-6.

30. Hirayama, M., Okugawa, R., Ishibashi, S., Murakami, S. & Miyake, T. Weyl node and spin texture in trigonal Tellurium and Selenium. *Phys. Rev. Lett.* **114**, 206401 (2015).

31. Luryi, S. Quantum capacitance devices. *Appl. Phys. Lett.* **52**, 501–503 (1988).

32. Ilani, S., Donev, L. A. K., Kindermann, M. & McEuen, P. L. Measurement of the quantum capacitance of interacting electrons in carbon nanotubes. *Nat. Phys.* **2**, 687–691 (2006).

33. Xia, J., Chen, F., Li, J. & Tao, N. Measurement of the quantum capacitance of graphene. *Nat. Nanotechnol.* **4**, 505–509 (2009).

34. Qiu, G. *et al.* Quantum Hall effect of Weyl fermions in n-type semiconducting

tellurene. *Nat. Nanotechnol.* **15**, 585–591 (2020).

35. Qiu, G. *et al.* Quantum transport and band structure evolution under high magnetic field in few-layer tellurene. *Nano Lett.* **18**, 5760-5767 (2018).

36. Niu, C. *et al.* High-pressure induced Weyl semimetal phase in 2D Tellurium. *Commun. Phys.* **6**, 345 (2023).

37. Tsirkin, S. S., Puente, P. A. & Souza, I. Gyrotropic effects in trigonal tellurium studied from first principles. *Phys. Rev. B* **97**, 035158 (2018).

38. Qiu, G. *et al.* High-performance few-layer tellurium CMOS devices enabled by atomic layer deposited dielectric doping technique. in *Device Research Conference - Conference Digest, DRC* (2018). doi:10.1109/DRC.2018.8442253.

39. Xu, K. *et al.* Influence of substrate impurity concentration on sub-threshold swing of Si n-channel MOSFETs at cryogenic temperatures down to 4 K. *Jpn. J. Appl. Phys.* **62**, SC1062 (2023).

40. Liu, X. *et al.* Fermi level pinning dependent 2D semiconductor devices: challenges and prospects. *Adv. Mater.* **34**, 2108425 (2022).

41. Dimoulas, A., Tsipas, P., Sotiropoulos, A. & Evangelou, E. K. Fermi-level pinning and charge neutrality level in germanium. *Appl. Phys. Lett.* **89**, 252110 (2006).

## Methods

**Hydrothermal growth of 2D Te flakes.** 1 g of polyvinylpyrrolidone (PVP) (Sigma-Aldrich) and 0.18 g of $Na_2TeO_3$ (Sigma-Aldrich) were dissolved in 64 ml double-distilled water. 6.66 ml of aqueous ammonia solution (25-28%, w/w%) and 3.34 ml of hydrazine hydrate (80%, w/w%) were added to the solution under magnetic stirring to form a homogeneous solution. The mixture was sealed in a 100 ml Teflon-lined stainless-steel autoclave and heated at 180 °C for 20 hours before naturally cooling down to room temperature.

**Device fabrication.** Te flakes were transferred onto 90 nm $SiO_2$/Si substrate or prepatterned 40 nm Ti substrate using Langmuir-Blodgett method. The 2D Te FETs were patterned using electron beam lithography and metal contacts (Ni) and gate dielectric (Ti and Al) were deposited by electron beam evaporation. Optional 3 nm atomic layer deposition (ALD) grown $Al_2O_3$ is used to tune the threshold voltage of the 2D Te FETs at 120 °C using $(CH_3)_3Al$ (TMA) and $H_2O$ as precursors.

**High resolution scanning transmission electron microscopy (HR-STEM).** TEM, selected area diffraction, energy dispersive x-ray spectroscopy (EDS) elemental mappings, and HAADF-STEM analysis were performed with FEI TALOS F200x. This microscope was operated with an acceleration voltage of 200 kV.

**Low-temperature electrical measurements.** The low temperature characterization was performed in a Lakeshore CRX-VF cryogenic probe station. The electrical characterization of transistors and inverters was measured with the Keysight B1500 system. The capacitance vs. voltage measurements were performed using Agilent E4980A LCR Meter at an excitation of 20 mV.

**Acknowledgements**

P.D.Y. was supported by Army Research Office under grant No. W911NF-15-1-0574 and by NSF under grant No. CMMI-1762698. Y.Z. and H.W. acknowledge the support from the US National Science Foundation for the microscopy work (DMR-2016453). W.W. was sponsored by the Army Research Office under Grant Number W911NF-20-1-0118.

## Author Contributions

P.D.Y. supervised the project. C.N. and L.L. designed the experiments. L.L., C.N., Z.L., and J.-Y. Lin fabricated the devices. L.L., C.N., P.T., and W.W. synthesized the materials. Y.Z. and H.W. carried out the TEM/STEM measurements and image analysis. C.N. and L.L. performed the electrical measurements and analyzed the data. P.D.Y., C.N. and L.L. wrote the manuscript and all the authors commented on it.

## Competing financial interests

The authors declare no competing financial interests.

## Supporting Information

Additional details for NiTe$_x$-Ni intimate contact resistivity, top-gate back-gate dependence of the transfer charactristics, Arrhenius plot, Schottky barrier extraction, butterfly curve, and noise margin extraction are in the supplementary information.

## Corresponding Author

* Peide D. Ye (E-mail: yep@purdue.edu)

# Figures

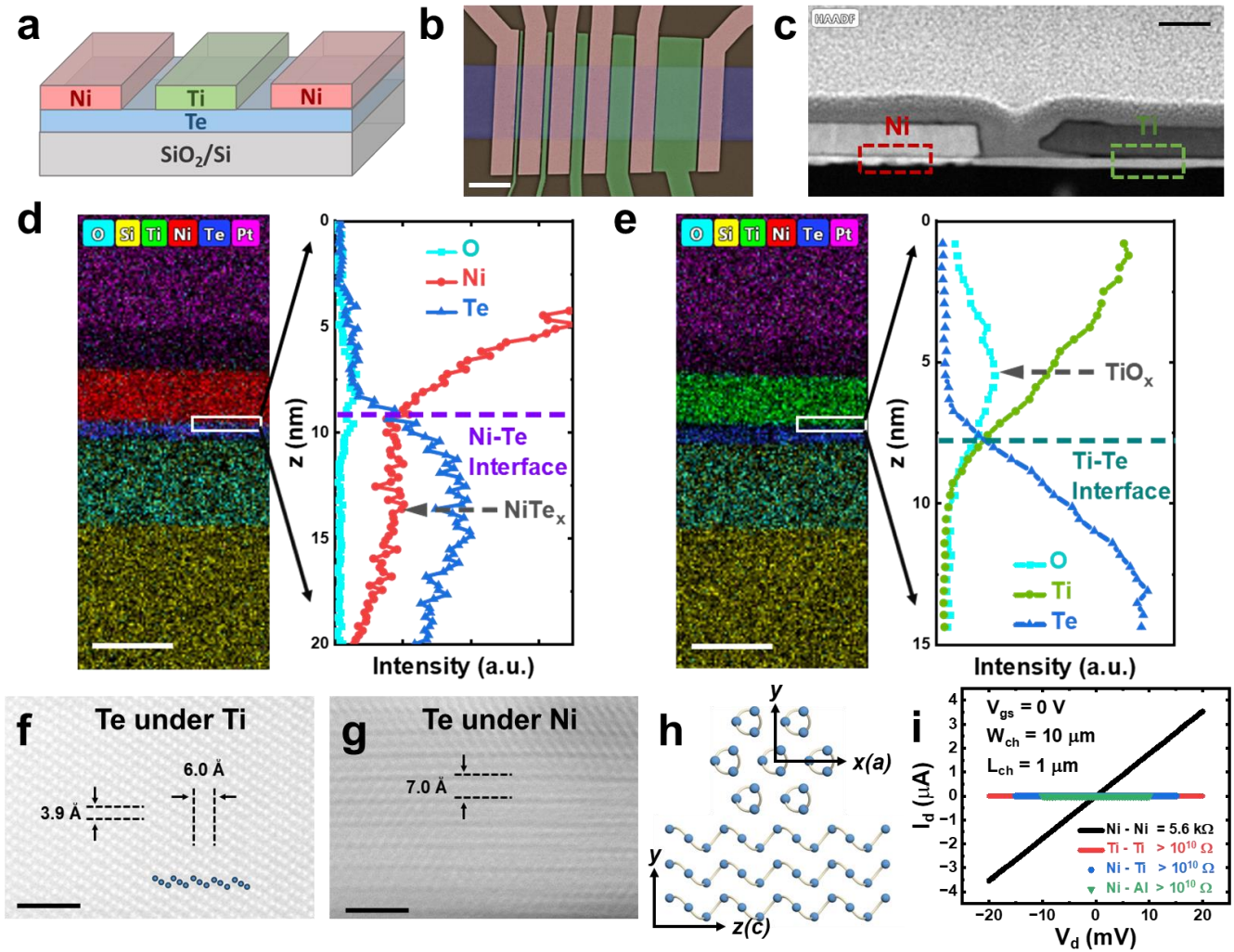

**Figure 1. Atomically Sharp Self-Formed Metal Contacts and Gate Dielectrics in 2D Te. a**, Schematic of a self-assembled 2D Te FET with directly deposited Ni as the contact and Ti as the gate. **b**, SEM image of a self-assembled 2D Te FET. The scale bar represents 2 μm. **c**, Cross-sectional HAADF-STEM image of the 2D Te FET. The scale bar represents 100 nm. **d**-**e**, EDS elemental mapping and line profiles at the Ni-Te (**d**) and Ti-Te (**e**) interfaces, indicating the formation of NiTe$_x$ metal contacts and TiO$_x$ gate dielectrics. The scale bar represents 100 nm. **f-g**, STEM images showing crystallized NiTe$_x$ at the Ti gate region (**f**) and the Ni contact region (**g**). The scale bar represents 2 nm. **h**, Schematic of the crystal structure of 2D Te. **i**, Electrical properties of different metal-to-Te electrodes. Ti and Al exhibit insulating behavior, making them suitable for gate electrodes.

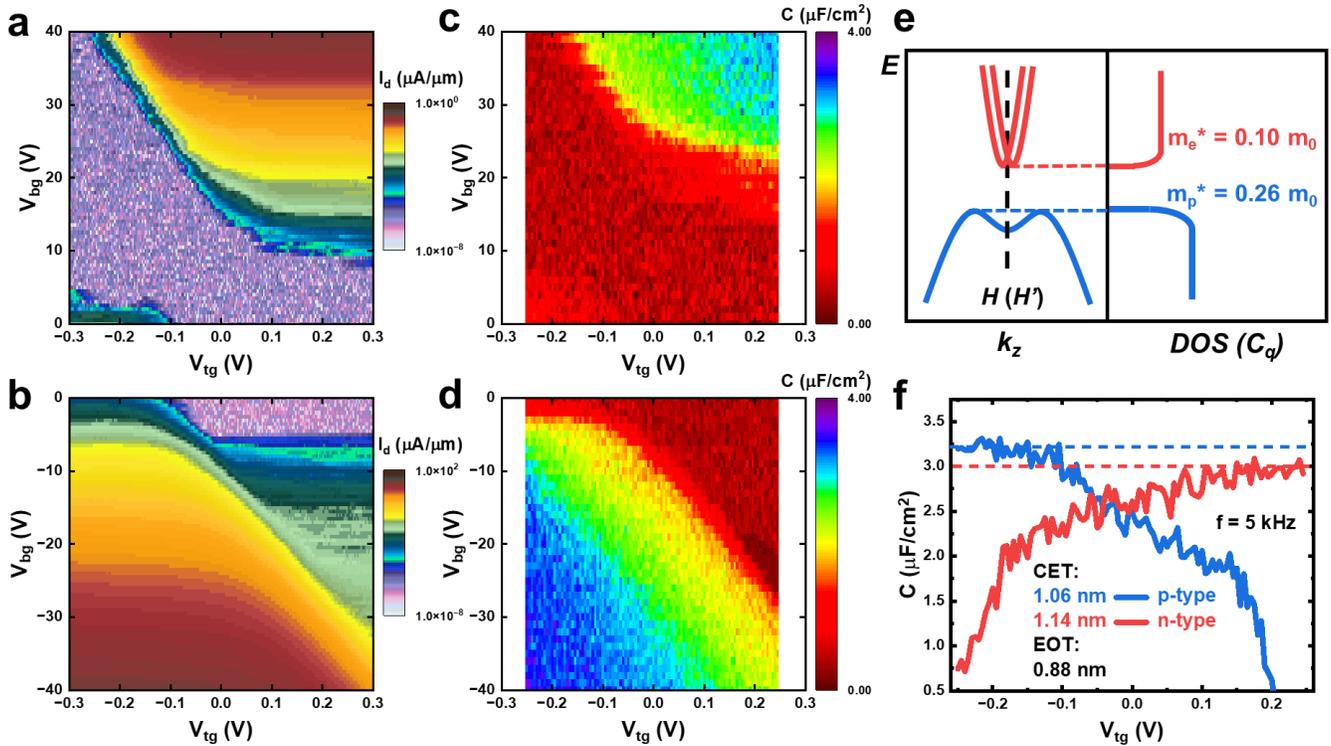

**Figure 2. Highly Efficient Modulation of Carriers (Low CET) and Effects of Quantum Capacitance. a-b**, Color maps of the drain current ($I_d$) as a function of back gate (90 nm SiO$_2$) voltage and top gate (self-assembled TiO$_x$) voltage, showing highly efficient modulation for both n-FET (**a**) and p-FET (**b**) in the same device. **c-d**, Color maps of the top gate capacitance (C) as a function of back gate and top gate bias. **e**, Schematic of Te band structure around the Fermi level. The differing effective masses at the conduction and valence bands result in distinct DOS and quantum capacitance ($C_q$). **f**, Capacitance vs. top gate voltage for n-type and p-type Te in the same device, with CET values of 1.06 nm (p-type) and 1.14 nm (n-type). The difference in capacitance is attributed to the effect of quantum capacitance. The EOT of the top gate is extracted to be 0.88 nm.

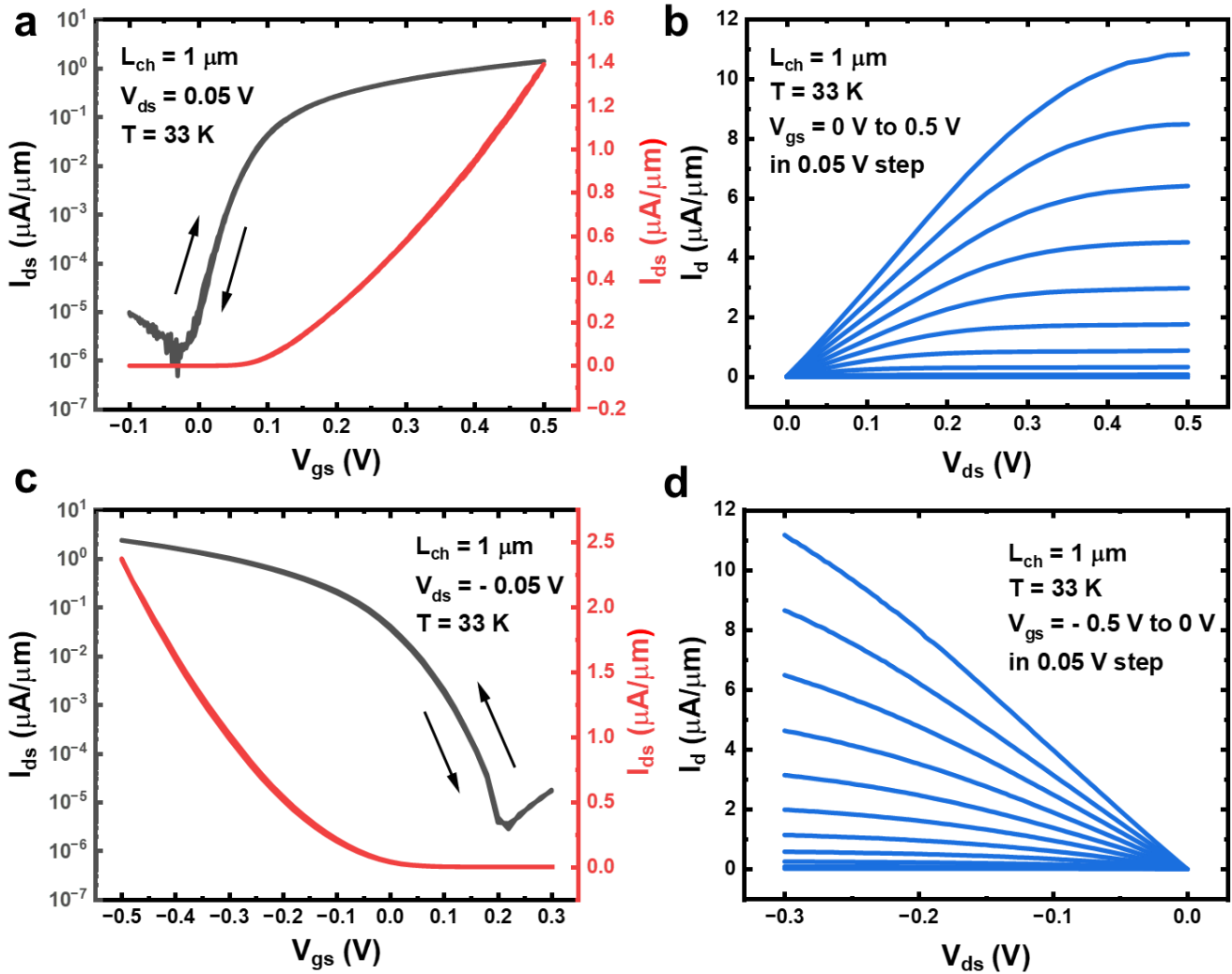

**Figure 3. Low Voltage Operation of 2D Te n- and p-FETs Using Ti as Back Gate. a**, Transfer characteristic of a 2D Te n-FET with a 1 μm channel length at 33 K. The n-type doping is achieved by depositing 3 nm $Al_2O_3$ via ALD on the 2D Te surface. **b**, Output characteristic of the same n-FET device shown in **a**, demonstrating good saturation behavior. **c**, Transfer characteristic of a 2D Te p-FET with a 1 μm channel length at 33 K. **d**, Output characteristic of the same p-FET device shown in **c**.

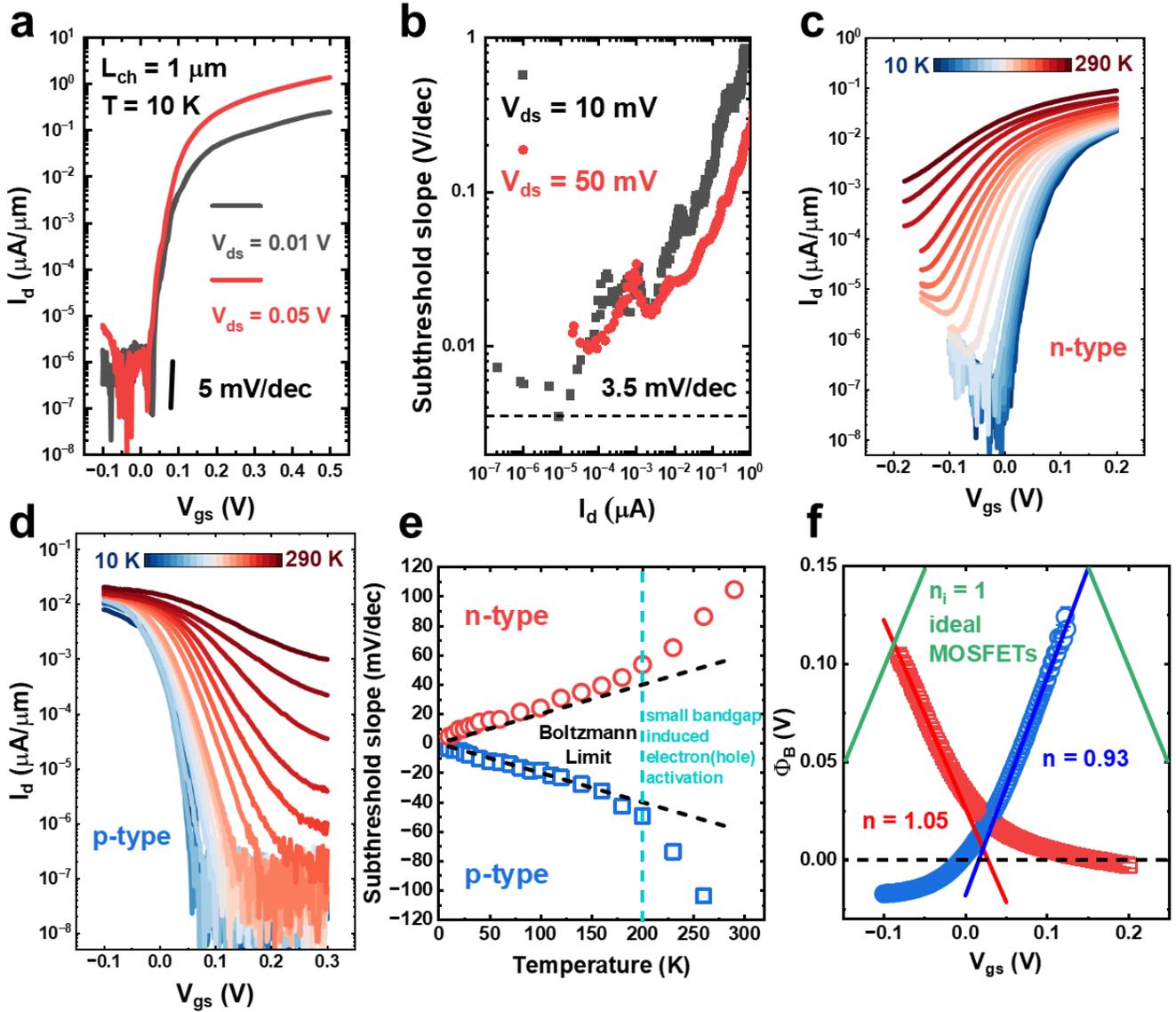

**Figure 4. Near-Ideal Subthreshold Slope (SS) in 2D Te n- and p-FETs Enabled by Ultraclean Self-Formed Semiconductor-to-Dielectric Interface.** **a**, Transfer characteristic of a 2D Te n-FET at 10 K under different $V_{ds}$. **b**, Subthreshold slope (SS) as a function of the drain current, demonstrating an ultralow SS of 3.5 mV/dec, indicative of the ultraclean interface. **c-d**, Temperature dependence of transfer characterizations for n-FET (**c**) and p-FET (**d**). **e**, SS of 2D Te n- and p-FET as a function of temperature. The SS closely follows the Boltzmann limit below 200 K, with deviations at higher temperatures attributed to the thermal activation of electrons in p-FETs and holes in n-FETs. **f**, Barrier

height as a function of gate voltage extracted from (**c**) and (**d**), demonstrating near-ideal gate tunability in both p-FET and n-FET.

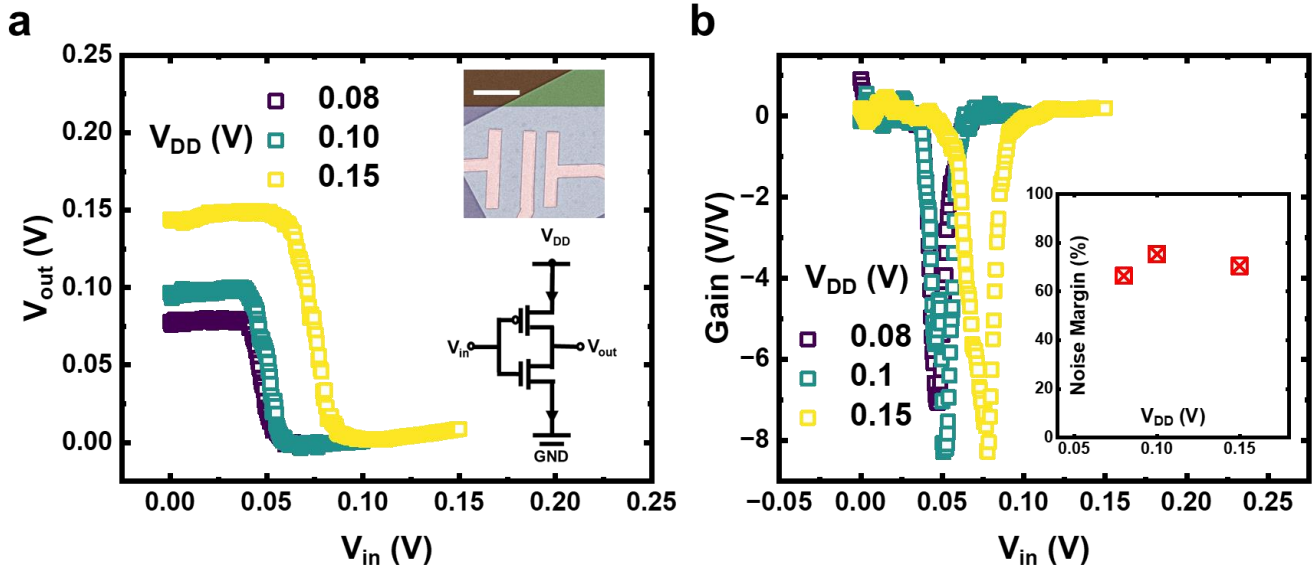

**Figure 5. Ultralow Voltage Operation of 2D Te CMOS Inverters. a**, Voltage transfer characteristic of a typical 2D Te CMOS inverter operating at an ultralow voltage of 0.08 V. Inset: SEM image and schematic of the 2D Te CMOS inverter. **b**, Voltage gains at different $V_{DD}$ extracted from **a**. A high gain of 7.1 V/V is achieved at 0.08 V $V_{DD}$. Inset: Nosie margin of the inverter at different $V_{DD}$. The measurement temperature is 10 K.

Supplementary Information for:

# Ultralow Voltage Operation of p- and n-FETs Enabled by Self-Formed Gate Dielectric and Metal Contacts on 2D Tellurium

Chang Niu[1,2,†], Linjia Long[1,2,†], Yizhi Zhang[3], Zehao Lin[1,2], Pukun Tan[1,2], Jian-Yu Lin[1,2], Wenzhuo Wu[4], Haiyan Wang[3] and Peide D. Ye[1,2,*]

[1]*Elmore Family School of Electrical and Computer Engineering, Purdue University, West Lafayette, IN 47907, United States.*

[2]*Birck Nanotechnology Center, Purdue University, West Lafayette, IN 47907, United States.*

[3]*School of Materials Science and Engineering, Purdue University, West Lafayette, Indiana 47907, United States.*

[4]*School of Industrial Engineering, Purdue University, West Lafayette, IN 47907, United States.*

†These authors contributed equally to this work: Chang Niu, Linjia Long

*Correspondence and requests for materials should be addressed to P. D. Y. (yep@purdue.edu)

**List of contents:**

**Supplementary figures:**

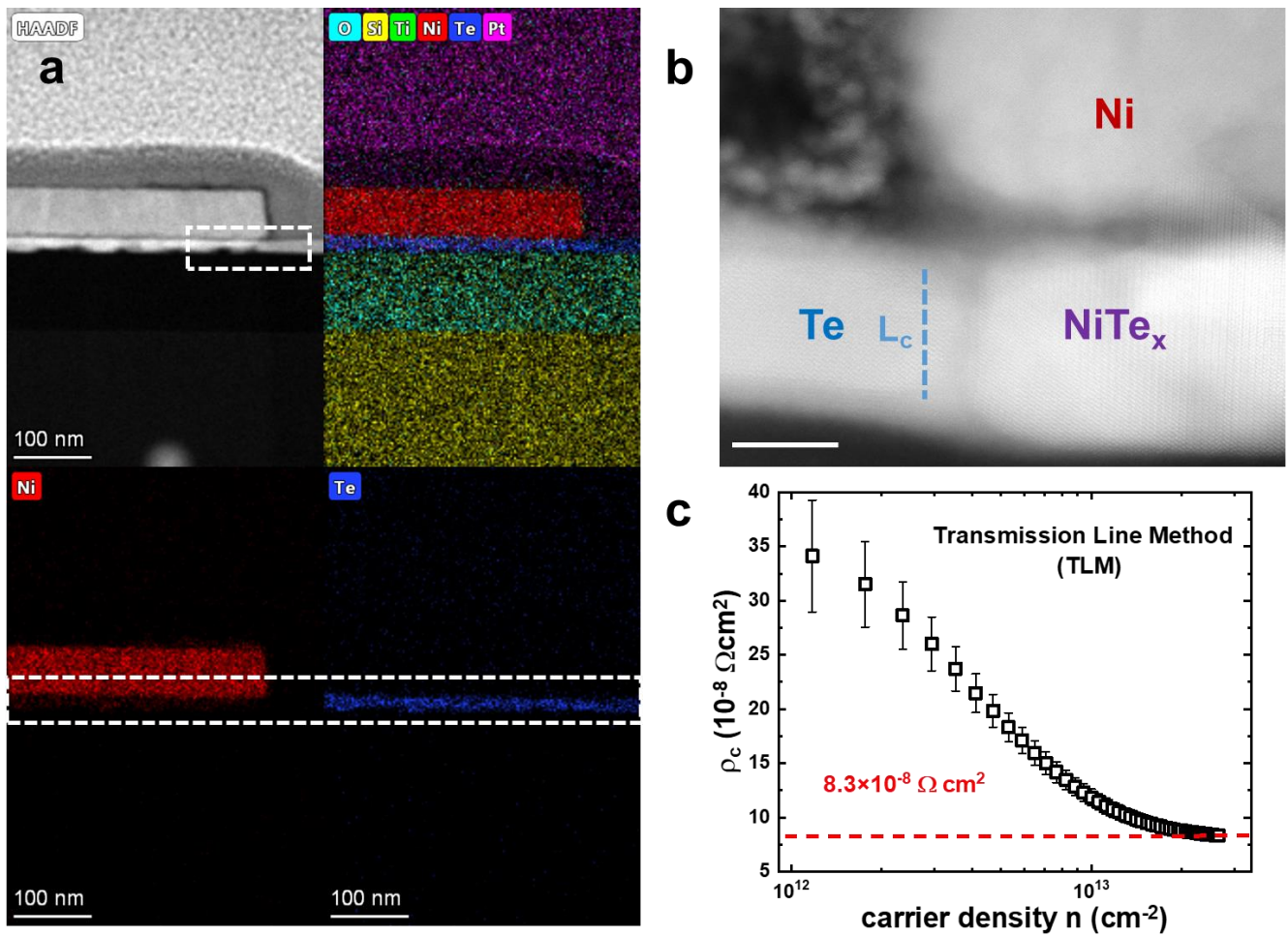

**Figure S1. Intimate NiTe$_x$-Ni contacts. a**, HAADF-STEM images, and EDS elemental mapping of the 2D Te devices under Ni contact. The contrast different between Te under Ni and Te channel is highlighted. The diffusion of Ni is observed in EDS mapping. **b**, STEM image of the contact to channel interface. The interface is atomically sharp. **c**, Contact resistivity at different carrier densities of a 2D Te p-FET extracted using the transmission line method (TLM).

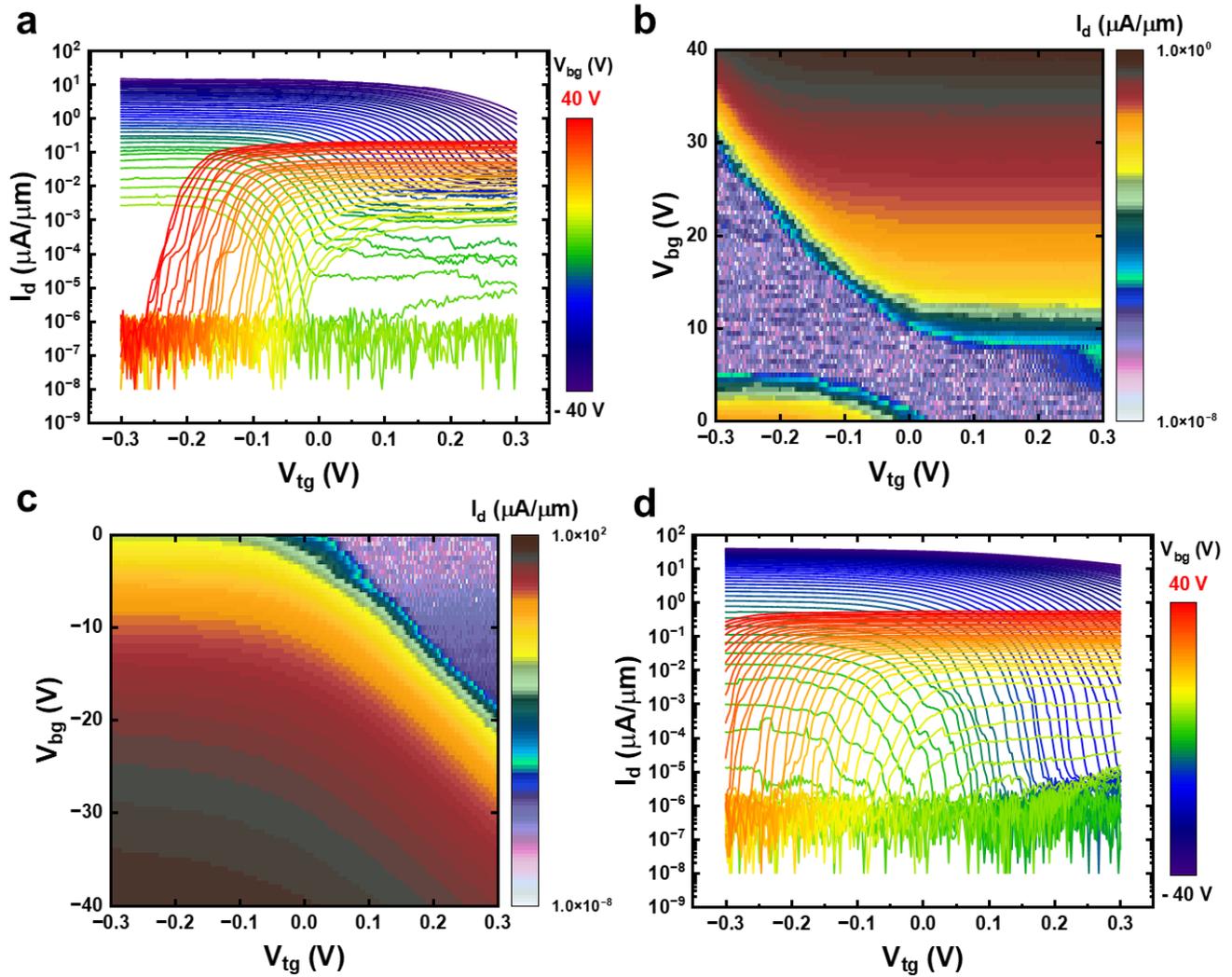

**Figure S2. Top-gate and back-gate dependence of the transfer characteristics. a**, Back gate voltage dependence of transfer characteristics of the same device in Figure 2a and 2b. **b-d**, Another similar 2D Te dual gate device with similar behavior, showing large top-gate tunability.

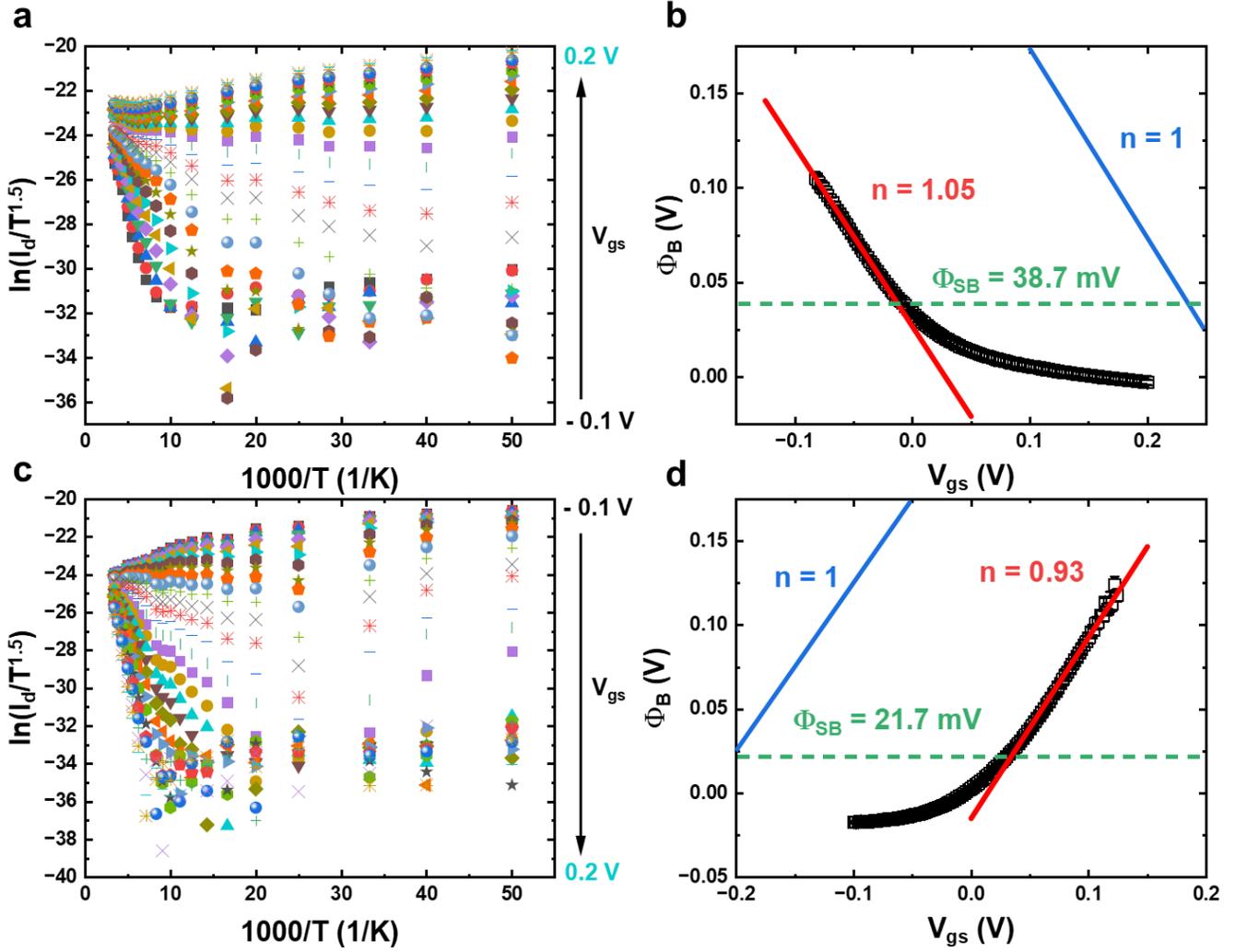

**Figure S3. Schottky barrier extraction for n-FET and p-FET. a**, Arrhenius plot at different gate voltages for an n-type 2D Te FET. **b**, Gate-voltage dependence of the barrier height extracted from the slope of the Arrhenius plot at high temperatures. The flat band condition is determined using linear fitting. A Schottky barrier of 38.7 mV is extracted. **c**, Arrhenius plot at different gate voltages for a p-type 2D Te FET. **d**, A Schottky barrier of 21.7 mV is extracted.

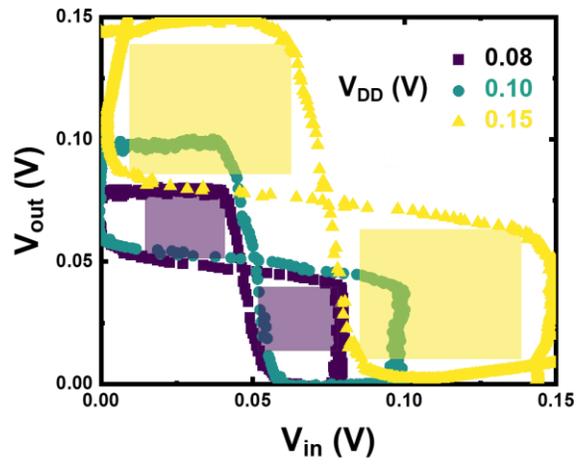

**Figure S4. Noise margin extraction of the CMOS inverter.** Butterfly curves under different supply voltages. The noise margin is extracted using the largest possible square method.